\documentclass[prl,twocolumn,amssymb,amsmath,aps,superscriptaddress,showpacs]{revtex4}
\usepackage{amsfonts}
\usepackage{graphicx}
\usepackage{epsfig}
\def\bea{\begin{eqnarray}}
\def\eea{\end{eqnarray}}
\def\be{\begin{equation}}
\def\ee{\end{equation}}

\begin{document}
\title{Nonequilibrium Transport through Double Quantum Dots: Exact Results near
Quantum Critical Point}
\author{Eran Sela and Ian Affleck \date{\today}}
\affiliation{Department of Physics and Astronomy, University of
British Columbia, Vancouver, B.C., Canada, V6T 1Z1}

\begin{abstract}
We study a double quantum dot in the regime where each dot carries a
spin-$1/2$. This system is described by the 2-impurity Kondo model,
having a non-Fermi liquid fixed point for a critical value of the
inter-impurity coupling. The Hamiltonian describing the vicinity of
the critical point, including the relevant potential scattering
perturbations, can be cast in quadratic form. This allows us to
predict a universal scaling function for the finite temperature
nonlinear conductance along the crossover from the critical point to
the surrounding stable fixed points.
\end{abstract}

\maketitle

\paragraph*{Introduction.} It is now well established that quantum dots (QD's) behave as Kondo
impurities at low temperatures~\cite{Gordon98,Cronenwett98}. So far,
most experiments focused on the study of Fermi-liquid states, with
regular thermodynamic as well as nonequilibrium transport
properties~\cite{Grobis08} at low temperatures. However, recently a
controlled QD setup was constructed experimentally to access the
two-channel Kondo fixed point~\cite{Potok07}. Remarkably, the
conductance of this fixed point matches a non-Fermi-liquid (NFL)
scaling function over a broad range of energy.

Another simple impurity model showing NFL behavior, similar to the
two-channel Kondo model, is the two-impurity Kondo model (2IKM),
consisting of two impurity spins that are coupled to conduction
electrons and, at the same time, interact with each other through an
exchange interaction $K$. Jones \emph{et. al.}~\cite{Jones88}
observed that in the 2IKM a NFL quantum critical point (QCP) at
$K=K_c$ separates a ''local singlet'' from a Kondo-screened phase.
The exact critical behavior was found using conformal field theory
methods in~\cite{Affleck92,Affleck95}. Georges and
Meir~\cite{Georges99} and Zar\'{a}nd \emph{et. al.}~\cite{Zarand06}
proposed  different QD realizations of this critical point.

\begin{figure}[h]
\begin{center}
\includegraphics*[width=60mm]{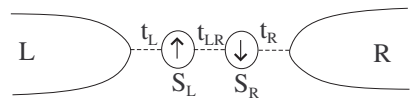}
\caption{Double quantum dot system. \label{fg:1}}
\end{center}
\end{figure}

In this paper we study a series double quantum dot realizing
the 2IKM when each dot behaves as an effective spin [see
Fig.~(\ref{fg:1})]. Ref.~\cite{Georges99} studied this system in a
wide range of parameters at zero temperature using a slave boson
mean field theory. Here we concentrate on the vicinity of the
critical point and predict exact \emph{full crossover} functions for
the nonlinear conductance from the QCP to the surrounding manifold
of fixed points at finite temperature and source drain voltage. This
crossover is calculated taking into account all three relevant
perturbations at the QCP. The first perturbation is associated with
$K-K_c$. Whereas typically potential scattering is a marginal
perturbation in quantum impurity models, in the 2IKM it leads to two
additional relevant operators at the critical
point~\cite{Affleck92,Affleck95,Zarand06}.

Our results are obtained using a convenient basis in which the QCP
is described by a quadratic Hamiltonian. This Hamiltonian follows
from Gan's theory for the 2IKM~\cite{Gan95}. While this theory was
derived for a spin anisotropic version of the 2IKM, it was argued
that it describes correctly also the spin SU(2) symmetric critical
point~\cite{Gan95}. We substantiate this statement further by
showing explicitly that the operators at the critical point have the
same form as in the conformal field theory~\cite{Affleck92}.

In our crossover calculation of the conductance we include not only
finite temperature, but also address the non-equilibrium problem at
finite source drain voltage. Originally this problem was addressed
using perturbation theory~\cite{Appelbaum66}, valid at energy scales
large compared to the Kondo temperature $T_K$. To address the
crossover to low energies different nonperturbative techniques were
adopted. Ref.~\cite{Schiller98} studied transport through a
1-channel Kondo impurity using abelian bosonization; however exact
results were obtained only for a specific point in the parameters
space (Toulouse limit). Another important development in this
direction was the application of the Bethe-ansatz and finding of
many-body scattering states~\cite{Konik02,Mehta06}.

\paragraph*{Model.}
Double QDs have already been proven to display rich Kondo physics in
competition with inter-dot exchange
interaction~\cite{Zarand06,Georges99,Lopez02,Jeong01,Craig04,Konic07}.
A multilevel single QD near a singlet triplet transition is also
described by the 2IKM~\cite{Pustilnik00}.

Our model consists of two leads, denoted L and R (or $i=1$ or $2$
respectively), attached to two QDs which behave as effective spins
$\vec{S}_L$, $\vec{S}_R$ due to large Coulomb energy $U$. After the
standard ``unfolding transformation"~\cite{Affleck91}, one obtains a
description of the model in terms of 4 left moving Dirac fermions
$\psi_{i \alpha}(x)$, $i=L,R$, $\alpha=\uparrow,\downarrow$, $x \in
\{-\infty,\infty \}$. We use a convention in which $\hbar=v_F=1$,
where $v_F$ is the Fermi velocity. At the moment let us assume
particle-hole symmetry. The interaction Hamiltonian generated to leading order
in the tunneling amplitudes $t_L$, $t_R$, $t_{LR}$ takes the form
\begin{eqnarray}
\label{eq:HK} H_K=J_L \vec{s}_{LL} \cdot \vec{S_L} +J_R \vec{s}_{RR}
\cdot \vec{S_R} +K \vec{S}_L \cdot \vec{S}_R,
\end{eqnarray}
with $\vec{s}_{ij} = {\psi^\dagger}^{ \alpha}_i
\frac{\vec{\sigma}_\alpha^\beta}{2} \psi_{j \beta}$, $J_i \sim
t_i^2/U$, and $K \sim t_{LR}^2/U$.

At $t_{LR}=0$ the Kondo effect takes place separately between the
left lead and left dot, and between the right lead and right dot.
The system is insulating in this case. This Kondo-screened phase
persists at positive nonzero $K$ up to $K=K_c$, above which a
''local singlet'' forms between the two
impurities~\cite{Jones88,Zarand06}. The critical value of $K$ occurs
at  $K_c \sim T_K $. Here we assume roughly equal Kondo temperatures
for left and right sides, $T_{KL}\approx T_{KR}\equiv T_K$, but most
of our results do not depend on that assumption, as discussed below.
The fixed point at $K=K_c$ has NFL properties and is unstable with
respect to an operator with anomalous dimension $\Delta =1/2$
associated with the perturbation $K-K_c$. However for the general
case when particle-hole and parity ($L \leftrightarrow R$)
symmetries are broken the QCP is unstable against two additional
relevant operators identified with potential
scattering~\cite{Affleck95,Zarand06}. It was found that while intra
lead potential scattering terms $V_{LL} \psi^\dagger_L \psi_L +
V_{RR} \psi^\dagger_R \psi_R $ are marginal at the QCP, inter-lead
direct tunneling terms
\begin{equation}
\label{eq:HPS} H_{PS} =  V_{LR} \psi^\dagger_L \psi_R +\emph{h.c.}
\end{equation}
correspond to two relevant $\Delta =1/2$ operators. In our model
$|V_{LR}| \sim t_L t_{LR} t_R/U^2$. When parity is broken $V_{LR}$
can be complex.

\paragraph*{Fixed point Hamiltonian.}
We bosonize the original theory and introduce four left moving
bosonic fields $\phi_{j \alpha }$ satisfying $\psi_{j \alpha} \sim
e^{i \phi_{j \alpha }}$. Subsequently 4 linear bosonic combinations
are defined, corresponding to charge, spin, flavor, and difference
of spin between the flavors: $ \phi_{c} =\frac{1}{2} \sum_{j \alpha}
\phi_{j \alpha}$, $\phi_{s} =\frac{1}{2} \sum_{j \alpha}
(\sigma^z)_\alpha^\alpha \phi_{j \alpha}$, $\phi_{f} =\frac{1}{2}
\sum_{j \alpha} (\tau^z)_j^j \phi_{j \alpha}$, and $\phi_{X}
=\frac{1}{2} \sum_{j \alpha} (\tau^z)_j^j
(\sigma^z)_\alpha^\alpha\phi_{j \alpha}$. Here $\vec{\sigma}$ and
$\vec{\tau}$ are Pauli matrices acting on the spin and $L/R$ space
respectively. Since the exponentials of the new bosons have
dimension $\Delta =1/2$, we can define new fermions $\psi_A=e^{i
\phi_A}$, $A=c,s,f,X$. Taking the real and imaginary parts of those
fermions we obtain 8 Majorana fermions
$\protect{\chi_1^A=\frac{{\psi^\dagger}^A+\psi_A}{\sqrt{2}}}$, $
\protect{\chi_2^A=\frac{{\psi^\dagger}^A-\psi_A}{\sqrt{2}i}}$. This
is denoted as the $SO(8)$ Majorana representation of the free
fermions, due to the $SO(8)$ symmetry group at the trivial fixed
point $J=0$.

At the trivial fixed point $J=0$ the free Majorana fermions have
trivial boundary condition (BC) $\chi_i^A(0^-)=\chi_i^A(0^+)$,
$i=1,2$, $A=c,s,f,X$, implying the continuity of those fields at the
boundary $x=0$. The fermions are still free at the critical point
and obey the BC established in Ref.~\cite{Affleck92}, using a
Bose-Ising representation for the free fermions. It was found that
the nontrivial BC occurs in the Ising sector of the theory. By
relating this representation with the $SO(8)$ Majorana
representation, it is found that the only nontrivial part of the BC
reads $\chi_2^X(0^-)=-\chi_2^X(0^+)$~\cite{Maldacena97,Gan95}. This
follows by identifying $\chi_2^X$ with the Majorana fermion of the
Ising model in~\cite{Affleck92}.

It is our main point in this section to argue that for energy scales
well below $T_K$ the corrections to the free Hamiltonian of the QCP
take a simple quadratic form in the $SO(8)$ representation:
\begin{eqnarray}
\label{eq:3} \delta H &=& i \sum_{j=1}^3 \lambda_j \chi_j(0)
a, \\
\lambda_1 =c_1 \frac{K-K_c}{\sqrt{T_K}},&&(\lambda_2 ,\lambda_3)=c_2
{\sqrt{T_K}} \nu \bigr({\rm{Re}} V_{LR},{\rm{Im}} V_{LR}\bigl),
\nonumber
\end{eqnarray}
where $\chi_1(x)=\chi_2^X(x) {\rm{sgn}}(x)$,
$(\chi_2,\chi_3)=(\chi_1^{f},\chi_2^{f})$, $c_1$ and $c_2$ are
constant factors of order 1, and $a$ is a local Majorana fermion
$a^2=1/2$.

At $ \lambda_2 =\lambda_3=0$, Eq.~(\ref{eq:3}) corresponds to the
correction to the fixed point hamiltonian of the particle-hole and
parity symmetric model, Eq.~(\ref{eq:HK}). Using conformal field
theory methods it was shown that the relevant operator in this case
is equivalent to a magnetic field acting on the boundary spin of a
quantum Ising chain. This is consistent with the $j=1$ term of
Eq.~(\ref{eq:3}), since one can identify $\chi_1(x)$ with the Ising
fermion~\cite{Maldacena97} and write the boundary spin of the Ising
model as $\sigma_B = i \chi_1(x=0) a$~\cite{Ghoshal94}. In this
analogy $\lambda_1$ corresponds to the magnetic field acting on the
boundary spin.

In addition we will clarify that the $j=1$ term of Eq.~(\ref{eq:3})
is consistent with Gan's theory for the anisotropic Toulouse limit
of the 2IKM~\cite{Gan95}. This theory uses the $SO(8)$ Majorana
representation, and the two impurity spins turn into a local fermion
$d$, where $\{d,d^\dagger \} =1$. Defining two Majorana fermions
$a=\frac{d-d^\dagger}{\sqrt{2} i}$ and $b =
\frac{d+d^\dagger}{\sqrt{2}}$, the boundary part of Gan's
Hamiltonian reads
\begin{equation}
\label{eq:Gan} \delta H_{G} = 2 i\sqrt{T_K} \chi_2^X(0) b-i(K-K_c)a
b.
\end{equation}
In general $\delta H_{G} \ne \delta H|_{\lambda_2=\lambda_3=0}$,
however we shall show that the two coincide for energy scales $\ll
T_K$. To see this suppose $K=K_c$ and consider a mode expansion
\begin{eqnarray}
\nonumber \chi_2^X(x) = \sum^\Lambda_k (\varphi_k(x) \psi_k+
h.c.),~~~ b= \sum^\Lambda_k (u_k \psi_k+  h.c.),
\end{eqnarray}
where $\{ \psi_k , \psi^\dagger_{k`}\}=\delta(k-k')$, $\{ \psi_k ,
\psi_{k`}\}=0$, and where initially we choose $\Lambda \gg T_K$ as
an ultraviolet cutoff. Solving Schr\"{o}dinger`s equation for the
wave functions $\varphi_k(x)$ and $u_k$ one finds $\varphi_k(x)
=e^{i k x}[ \theta(x) \varphi^{(+)}_k+ \theta(-x) \varphi^{(-)}_k]$,
$\varphi_k(0) =\frac{1}{2}( \varphi^{(+)}_k+ \varphi^{(-)}_k)$, $u_k
= \frac{2}{i k}\sqrt{T_K}\varphi_k(0)$,
$\varphi^{(-)}_k/\varphi^{(+)}_k = e^{ 2 i \delta}$, $\tan
\delta=\frac{2 T_K}{ k}$. One should normalize $\varphi^{(+)}_k =
\frac{1}{\sqrt{\ell}}$, where $\ell$ is the size of the system.
While at $T_K=0$ we have the BC $\chi_2^X(0^+) = \chi_2^X(0^-)$, we
see from the wave function that the effect of the first term in
$\delta H_G$ is to modify this BC to $\chi_2^X(0^+) =-
\chi_2^X(0^-)$ for energies $\ll T_K$. The key observation is that
the following operator identity holds if one restricts the mode
expansion of its LHS and RHS to energies below a cutoff $\Lambda \ll
T_K$,
\begin{equation}
\label{eq:b} b  =\frac{1}{\sqrt{ T_K}} \chi_1(0),~~~\chi_1(x) =
\chi_2^X(x) {\rm{sgn}}(x).
\end{equation}
Physically this means that at energy scales below $T_K$ the local
operator $b$ is absorbed into the field $\chi_2^X$ and changes its
BC. Using the operator identity Eq.~(\ref{eq:b}), we see that the
term $\propto K-K_c$ in $\delta H_G$ is equivalent to the $j=1$ term
in $\delta H$. This establishes the connection between Gan's theory
and the boundary Ising model arising from the conformal field theory
solution, showing that Gan's anisotropic theory describes correctly
also the vicinity of the isotropic fixed point.

When particle-hole and parity symmetries are broken Gan's theory
predicts two additional relevant $\Delta =1/2$ operators which take
the form of the $j=2,3$ terms of Eq.~(\ref{eq:3}). The presence of
these two terms is consistent with the conformal field theory:
Consider the relevant potential scattering terms $\psi^\dagger_L
\psi_R + \psi^\dagger_R \psi_L$ and $i(\psi^\dagger_L \psi_R
-\psi^\dagger_R \psi_L)$ in Eq.~(\ref{eq:HPS}), which have the
$SO(8)$ representation $i \chi_2^X \chi_1^{f}$ and $i \chi_2^X
\chi_2^{f}$ respectively. We are interested in the form these
operators take near the NFL critical point. This can be obtained,
using conformal field theory methods, by applying double fusion in
the Ising sector. This operation acts on $\chi_2^X$ only, and
transforms it into the identity operator. To maintain the bosonic
nature of the Hamiltonian a local dimension 0 Majorana fermion must
replace $\chi_2^X$. To account for the correct ground state
degeneracy of the QCP one requires this local Majorana fermion to
coincide with the same operator $a$ which couples to $\chi_2^X$ at
$K \ne K_c$. Therefore the two relevant potential scattering terms
must have the form of the $j=2,3$ terms of $H_{QCP}$.

\paragraph*{Conductance.}
With the fixed point Hamiltonian Eq.~(\ref{eq:3}) we proceed to
calculate the conductance of the double dot along the crossover from
the critical point to the surrounding fixed points, as function of
temperature $T$, source drain voltage $V$, and the three coupling
constants $\lambda_1$, $\lambda_2$, and $\lambda_3$. As we shall
see, the crossover is controlled by an energy scale
\begin{equation}
\label{eq:Tstar} T^* = \lambda_1^2+\lambda_2^2+\lambda_3^2. \end{equation}
Our
result, Eq.~(\ref{eq:conV}), is valid for any $T/T^*$ and $eV/T^*$,
as long as $eV, T$,
 $T^* \ll \min \{T_{KL},T_{KR} \}$.

\begin{figure}[h]
\begin{center}
\includegraphics*[width=60mm]{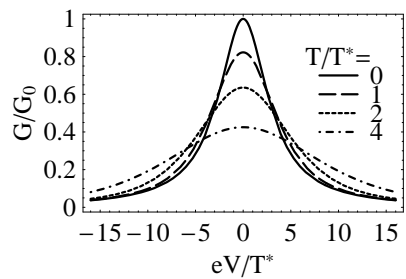}
\caption{Scaling function for the nonlinear conductance.
\label{fg:Ftv}}
\end{center}
\end{figure}

Schiller and Hershfield~\cite{Schiller98} studied the related
problem of transport through a single QD with effective spin
$\vec{S}$. They considered the anisotropic Kondo interaction of the
form $H_K^{1D} = \sum_{i,j=L,R} \sum_{\eta=x,y,z} J_\eta^{ij}
s_{ij}^\eta S^\eta$ with $J_x^{ij}=J_y^{ij}=J_\perp^{ij}$. Using the
$SO(8)$ representation, they obtained a quadratic theory [Eq.~(3.15)
at zero magnetic field $B=0$] for the case $J_z^{LR}=0$,
$J_z^{LL}=J_z^{RR}=2 \pi$, with a boundary term
\begin{equation}
\label{eq:H1D} \delta H_{1D} = \frac{i}{\sqrt{2 \pi \alpha}}
\bigl[J^+ \chi_X^2 b+J_\perp^{LR} \chi_f^1 a +J^- \chi_X^1 a \bigr],
\end{equation}
where $J^\pm=(J_\perp^{LL} \pm  J_\perp^{RR})/2$ and $\alpha^{-1}$
is an ultraviolet momentum cutoff. Comparing $\delta H_{1D}$ to
$\delta H_G$ in Eq.~(\ref{eq:Gan}), the coupling $J^+ \chi_X^2 b$ is
the only term involving $b$. For either the double or single QD the
current operator is given by $I = -i [H,Y]$ where $Y = \frac{1}{2}
\int_{-\infty}^\infty dx \bigl( {\psi^\alpha}^\dagger_L \psi_{L
\alpha} -{\psi^\alpha}^\dagger_R \psi_{R \alpha} \bigr)
=\int_{-\infty}^\infty dx \psi^\dagger_{f} \psi_{f}$. We see that
the current operator involves the flavor fermions, and hence the
first term $\propto J^+$ in Eq.~(\ref{eq:H1D}) does not contribute
to the conductance, since the flavor fermions are decoupled from
$b$. The operator $a$ in Eq.~(\ref{eq:H1D}) couples linearly to bulk
Majorana fields at $x=0$. Comparing to Eq.~(\ref{eq:3}), it is easy
to see that under the replacements $ \frac{J^{-}}{\sqrt{2 \pi
a}}\leftrightarrow \lambda_1$ and $\frac{J_{\perp}^{LR}}{\sqrt{2 \pi
a}} \leftrightarrow \sqrt{\lambda_2^2 + \lambda_3^2}$, the two
systems have the same current as function of temperature and
voltage. We shall not repeat the calculation of~\cite{Schiller98}
and present the result for the nonlinear conductance, $G = dI/dV$,
in terms of the parameters of our model
\begin{eqnarray}
\label{eq:conV} G = G_0 F \left[ \frac{T}{T^*}, \frac{eV}{T^*}
\right],\qquad G_0=\frac{2e^2}{h} \frac{T^*_{LR}}{T^*}
 ,  \nonumber \\
F[t,v]=\frac{1}{2t} {\rm{Re}} ~\psi'
\left(\frac{1}{2}+\frac{1}{2t}+\frac{i v}{2 \pi t}\right),
\end{eqnarray}
where $T^*_{LR}=\lambda_2^2 + \lambda_3^2$ and $\psi(z)$ is the
digamma function. We see that the energy scale $T^*$ defined in
Eq.~(\ref{eq:Tstar}) determines the crossover scale. At $K=K_c$,
$T^* \rightarrow T^*_{LR}$. The fact that $T^*$ is quadratic in
$K-K_c$ and in $V_{LR}$ is a signature of a NFL; these coupling
constants have fractional renormalization group scaling dimensions
$1/2$. In Fig.~(\ref{fg:Ftv}) we plot the conductance as function of
$eV/T^*$ for various values of $T/T^*$. At $T=0$,
$G/{G_0}=\bigl(1+(eV/\pi T^*)^2\bigr)^{-1}$ and for $V=0,T\ll T^*$,
$G/{G_0} \rightarrow 1-T^2/3 {{T^*}}^2$.

Note that the current vanishes when $V_{LR}=0$. This occurs since
the corresponding operators in Eq.~(\ref{eq:HPS}) are the only terms
in our model which do not conserve $Y$. At the fixed point we expect
additional terms not conserving $Y$, such as the marginal operators
$\chi_{1,2}^{f} \chi_2^X$. However the amplitude for such terms
transferring charge between the leads is small, and their effect can
be neglected.

Experimentally it is convenient to tune $K$ through $K_c$ by varying
$t_{LR}$, and observe the evolution of a conductance peak as
function of energy scales $T$ and $eV$. At $K=K_c$ there is a
residual energy scale
 $T^*_{LR}$ due to potential scattering. Using $|V_{LR}| \sim \frac{t_L t_{LR}
t_R}{U^2}$ and $K \sim \frac{t_{LR}^2}{U}\sim T_K$ we see that for a
typical QD the energy scale $T^*_{LR}$ is in the range of validity
of our theory, $T^*_{LR} \sim (\nu J)^2 T_K^2/U \ll
T_K$~\cite{Remark}. At low energies $T$, $eV \ll T^*$, we have $F
\rightarrow 1$ and $G \rightarrow G_0$. Thus $G$ exhibits a peak in
the low $T$ linear conductance with height $2e^2/h$ at $K=K_c$, and
width in $K-K_c$ which scales with $V_{LR}$.
 Such a peak structure was discussed in~\cite{Georges99};
here we give its precise form. Intra-lead potential scattering
$V_{LL}$ and $V_{RR}$ are marginal perturbations; they reduce the
conductance at the peak to
$G_{peak}=\frac{2e^2}{h}[1-\mathcal{O}(\nu J)^2 ]$. Now consider the
regime $T$, $eV \gg T^*$. In this case as $t_{LR}$ is varied the
system passes through the critical region at $K \sim K_c$ where the
system does not flow away from the NFL fixed point due to any
 of the relevant perturbations. The conductance now becomes:
\begin{equation} \label{eq:hight}
G \to  \frac{e^2}{h} \frac{T^*}{T}
 {\rm{Re}}~\psi'\left(\frac{1}{2}+\frac{ieV}{2\pi T}\right)=\frac{e^2}{h} \frac{T^*}{T} \frac{\pi^2}{2 \cosh^2 \frac{eV}{2 T}}.
\end{equation}
The width of the peak in $G$ as a function of source-drain voltage,
$V$, now scales with $T$.

Although we assumed $T_{KL}\approx T_{KR}$ above, in fact, most of
our results don't depend on that assumption since we have already
taken into account all relevant and marginal operators allowed in
the effective Hamiltonian when parity is broken. Only the precise
values of $T^*$, $T^*_{LR}$ and $K_c$ change in the asymmetric case,
and the necessary conditions to approach the critical region becomes
$T^*$, $T$, $eV\ll \min \{ T_{KR}, T_{KL} \}$. In particular, note
that the $T=0$ linear conductance has the value $2e^2/h$, for
$K=K_c$, independent of asymmetry in that limit. This is in striking
contrast to the $T=0$ linear conductance through a single Kondo
impurity which is suppressed by a factor of $2 |t_L
t_R|/(|t_L|^2+|t_R|^2)$. In the strongly asymmetric case with
$T_{KR} \ll T_{KL}$,  we find, up to logarithmic corrections, that
\begin{equation}
T^* \approx (K-K_c)^2 T_{KR}/T_{KL}^2+|\nu V_{LR}|^2T_{KR},
\end{equation}
$T^*_{LR} \approx |\nu V_{LR}|^2T_{KR}$, and $K_c \approx
T_{KL}$~\cite{Zarand06}.

As pointed out by Schiller and Hershfield~\cite{Schiller98}, their
single quantum dot model has the peculiarity of being related to the
two-channel Kondo NFL fixed point. This may be largely a consequence
of fine tuning to the Toulouse limit $J_z^{LR}=0$,
$J_z^{LL}=J_z^{RR}=2 \pi$. The two-channel and the two-impurity NFL
critical points are closely related. The double dot  system
discussed here provides a method of getting this NFL behavior in an
experimentally feasible way rather than by the unrealistic fine tuning
of parameters.

Our system differs from the proposal of Zar\'{a}nd \emph{et.
al.}~\cite{Zarand06}, in the way in which the leads couple to the
interacting system. In~\cite{Zarand06} only one linear combination
of the source and drain leads acts as a Kondo-screening channel,
whereas the second combination acts like a scanning tunneling
microscope tip. A third lead is needed as a second Kondo screening
channel. On the other hand the standard double dot system considered
here has only two leads and can be therefore easier to construct
experimentally. In~\cite{Zarand06} exotic temperature dependence for
the conductance $\sqrt{T}$ arises from the irrelevant operator.
However we find that the (same) irrelevant operator gives regular
quadratic corrections in $eV$ and in $T$ for our system~\cite{Sela}.
On the other hand the NFL signatures are apparent from the relevant
operators with dimension $1/2$ which can be addressed analytically
in our configuration.

\paragraph*{Conclusion.} In this paper we found new exact results for the 2IKM in
a double QD. We derived full crossover formulas for the conductance
from the NFL critical point to the stable fixed points in a
3-dimensional parameter space including $K-K_c$ and the two
additional relevant potential scattering operators. We clarified
that this crossover is described by a quadratic Hamiltonian.

We thank Amnon Aharony, Ora Entin-Wohlman, and Justin Malecki for
very helpful discussions. This work was supported by NSERC (ES $\&$
IA) and CIfAR (IA).


\begin{thebibliography}{99}
\bibitem{Remark} One can reduce
$T^*_{LR}$ further by constructing an ``antiferromagnetic insulator"
as proposed in~\cite{Zarand06}, by inserting two additional quantum
dots between $S_1$ and $S_2$.
\end{thebibliography}
\end{document}